%% file: final.tex
\documentclass{elsart}
\usepackage{graphics}
\usepackage{lscape}
\begin{document}
\begin{frontmatter}
\title{Measurement of the \nuc{6}{Li}(e,e$'$p) reaction cross sections 
at low momentum transfer}
\author[LNS]{T. Hotta\thanksref{rcnp}},
\author[LNS]{T. Tamae},
\author[LNS]{T. Miura},
\author[sci]{H. Miyase}, 
\author[LNS]{I. Nakagawa},
\author[sci]{T. Suda},
\author[LNS]{M. Sugawara},
\author[LNS]{T. Tadokoro},
\author[LNS]{A. Takahashi},
\author[LNS]{E. Tanaka},
\author[sci]{H. Tsubota}
\address[LNS]{Laboratory of Nuclear Science, 
Tohoku University, Mikamine, Taihaku, Sendai, Miyagi 982-0826, Japan}
\address[sci]{Physics Department, Graduate School of Science, Tohoku 
University, Aramaki, Aoba, Sendai, Miyagi 980-8578, Japan}
\thanks[rcnp]{Corresponding author.
Present address: Research Center for Nuclear Physics,
Osaka University, 10-1 Mihogaoka, Ibaraki, Osaka, 567-0047, Japan.
e-mail: hotta@rcnp.osaka-u.ac.jp
}
\begin{abstract}
The triple differential cross sections for the
\nuc{6}{Li}(e,e$'$p) reaction have
been measured in the excitation energy range from 27
to 46~MeV in a search for evidence of the giant 
dipole resonance (GDR) in \nuc{6}{Li}.
The cross sections have no distinct structures in this energy region,
and decrease smoothly with the energy transfer.
Angular distributions are different from those
expected with the GDR.
Protons are emitted strongly in the momentum-transfer direction.
The data are well reproduced by a DWIA calculation assuming a direct 
proton knockout process.
\end{abstract}
\end{frontmatter}
\tabcolsep=3.5pt
\section{Introduction}
The energy and width of the giant dipole resonance (GDR) smoothly
vary with the atomic number in medium and heavy nuclei~\cite{spet81},
while in light nuclei, the GDR is characterized
by specific features for each nucleus~\cite{era86}.
For 1p-shell nuclei, some theoretical studies predict that
the presence of the cluster structure causes
supermultiplet splitting of the GDR~\cite{era86}.
For \nuc{6}{Li}, Kurdyumov {\it et al.}~\cite{kur70} predicted by a translation
invariant shell model calculation that the GDR splits into three
multiplets with excitation energies of 10--12 MeV, 16--25 MeV
and 31--35 MeV.
According to their calculation, the state with the lowest excitation
energy decays by
proton or neutron emission leaving the residual nuclei,
\nuc{5}{He} or \nuc{5}{Li}, at the lowest $3/2^{-}$ and $1/2^{-}$
states.  
The state at 16--25~MeV decays through \nuc{3}{H}
emission in addition to the one-nucleon decay to the lowest states
of the residual nuclei.
The state at 31--35~MeV predominantly decays into
the $3/2^{+}$ state of the residual nuclei by emitting
a proton or a neutron,
whereas the decay into the lowest $3/2^{-}$ and $1/2^{-}$ states 
is forbidden because of 
spatial symmetry~\cite{kur70,fil86}.

The experimental situation, however, is still unsettled.
Using the yield-curve method,
Denisov {\it et al.}~\cite{den67} measured the photoproduction cross sections
of charged particles for \nuc{6}{Li} in the photon energy $E_{\gamma}$ 
region up to 55~MeV, and obtained the total photodisintegration cross 
section.
Although the cross section was obtained from rather small number of 
data points,
they claimed that three distinct peaks with a width about 5~MeV at about
11 MeV, 21 MeV and 30 MeV were observed
This result was consistent with the theoretical prediction.
However,
no later experiment has confirmed the presence of 
the GDR in \nuc{6}{Li} with such structures up to 
30~MeV~\cite{jun79,ber75,ahr75}.

In order to discuss the presence of the GDR excitation 
and its decay properties in \nuc{6}{Li},
we have measured the \nuc{6}{Li}(e,e$'$p) reaction cross sections 
for specific final states in the  
energy transfer region $\omega=27$--46~MeV, covering a sufficiently 
high-energy region where the highest component of the GDR is
predicted.
In the case of (e,e$'$p) reaction on heavier nuclei such as
\nuc{12}{C} and \nuc{40}{Ca}~\cite{cal94,neu94}, the GDR states are clearly 
observed as peaks in the cross sections at the same resonance energies 
observed in the photoreactions.
The angular distributions for decay protons
have the dipole characteristics, i.e. protons are strongly emitted to
both the parallel and anti-parallel directions to the momentum transfer
($\,\vec{q}\,$). 
  
In the high $\omega$ and $|\,\vec{q}\,|$ region, where the contribution from 
the giant resonances is small, the (e,e$'$p) reactions are well understood
as the direct knockout process, and a distorted wave impulse approximation
(DWIA) gives a good description of the reaction~\cite{lan89}.
In this case, protons are strongly emitted to the momentum transfer
direction reflecting the momentum distribution of the proton in the
ground state of the target nuclei.

It is expected that the contributions of the different reaction 
processes are identified by the angular distribution and its $\omega$
and $|\,\vec{q}\,|$ dependence.
Therefore, we have measured the angular distribution of emitted protons
at two different electron scattering angles: (1) $\theta_{\rm e}=26^{\circ}$ 
where the momentum transfer $|\,\vec{q}\,|$ is 60--67~MeV/$c$,
and (2) $\theta_{\rm e}=42^{\circ}$ where $|\,\vec{q}\,|$=90~MeV/$c$.

\section{Experimental Setup}
\begin{figure}
\begin{center}
\includegraphics{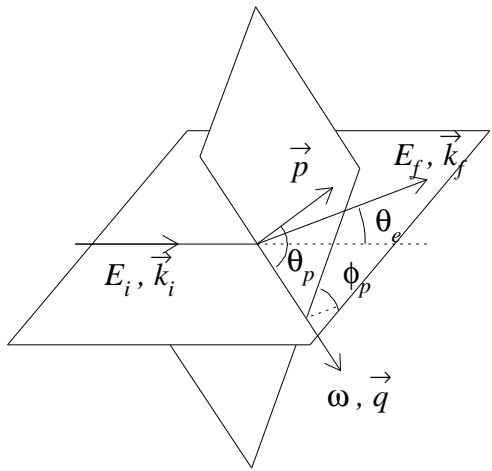}
\caption{Definition of the polar angle $\theta_{\rm p}$ and azimuthal
angle $\phi_{\rm p}$ of proton momentum $\vec{p}$ with respect to the
momentum transfer $\vec{q}$.
The incident- and scattered-electron momenta are 
represented by $\vec{k}_{\rm i}$ and $\vec{k}_{\rm f}$.}
\label{fig:kinematics}
\end{center}
\end{figure}

Figure~\ref{fig:kinematics} shows the kinematics of the (e,e$'$p) reaction
and the definition of reaction angles.
Energies of incident and scattered electrons are $E_{\rm i}$ and
$E_{\rm f}$, and $\theta_{\rm e}$ denotes the scattered electron angle.
The incident and scattered electrons define the scattering plane.
An energy $\omega=E_{\rm i}
-E_{\rm f}$ and a momentum $\vec{q}=\vec{k_{\rm i}}-
\vec{k_{\rm f}}$ are transferred to the target nuclei.
The direction of an emitted proton is defined by a polar angle 
$\theta_{\rm p}$ and an azimuthal angle $\phi_{\rm p}$
relative to the momentum transfer vector $\vec{q}$.

The experiment was carried out by using a 134 MeV continuous electron 
beam from the pulse stretcher ring, SSTR~\cite{SSTR} at 
the Laboratory of Nuclear Science, Tohoku University. 
The beam intensity was in the range of 150--300~nA.
The experimental setup is described in detail in a previous paper~\cite{tad94}.

We used a 95\% enriched, 6~mg/cm$^2$-thick \nuc{6}{Li} target.
Oxygen contamination in the target
was estimated from the elastic peak to be less than 0.1\%.
Scattered electrons were momentum-analyzed in a 
double-focusing magnetic 
spectrometer~\cite{LDM} having a solid angle of 2.9 msr, 
and detected in two layers of plastic scintillators
and a vertical drift chamber
(VDC) located in the focal plane.
A typical momentum resolution was about 0.2\% at 100~MeV/$c$.  
The electron spectrometer was set at 
$\theta_{\rm e}=26^{\circ}$ for 27~MeV$\le \omega \le$46~MeV,
and at $\theta_{\rm e}=42^{\circ}$ for 34~MeV$\le \omega \le $39~MeV.

Protons were measured with
detector telescopes, each consisting of one 50~$\mu$m-thick 
and 2 or 3 layers of 1~mm or 
2~mm-thick silicon surface barrier detectors (SSD)
with a sensitive area of 300~mm$^{2}$.
These telescopes were located 
at 12--18~cm from the target in every $30^{\circ}$ step.
They covered the angular range
$0^{\circ} \le \theta_{\rm p} \le 180^{\circ}$ for two planes having
$\phi_{\rm p}=-45^{\circ}$ and $-135^{\circ}$.
In order to reduce background, the detectors were shielded by
lead. A pair of permanent magnets in front of 
each telescope removed low energy electrons emitted from the target.
The acceptance solid angle of the telescope was defined 
by an iron collimator
2~mm in thickness, with an opening of 15~mm in diameter.
The number of incident electrons was assessed by using a secondary
emission monitor placed downstream of the target.

\section{Data Reconstruction}

\begin{figure}
\begin{center}
\includegraphics{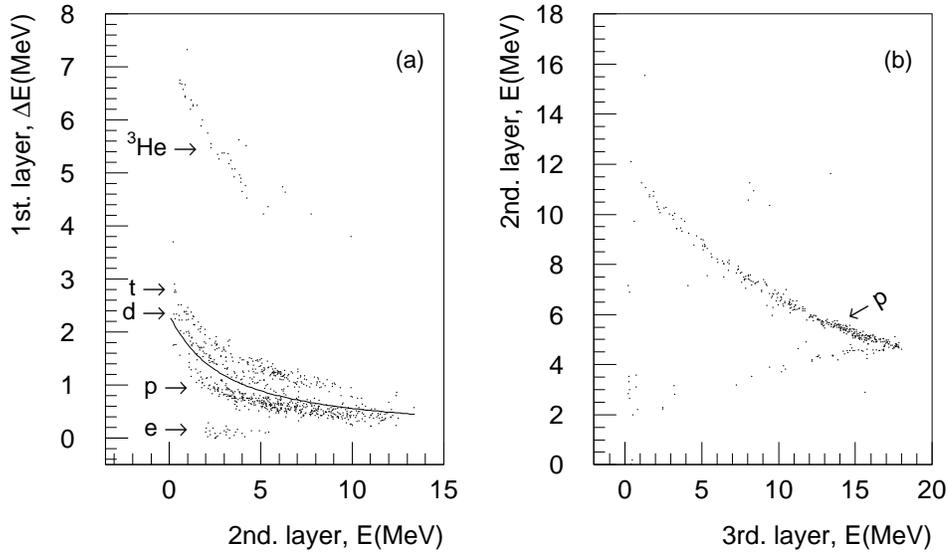}
\caption{Pulse height distributions for two layers of the SSD stack 
for particles stopping in (a) the second layer, and (b) in the third layer.}
\label{fig:2dadc}
\end{center}
\end{figure}

Charged particles reaching at least the second layer of 
the SSD stack were used for the analysis.
Particles stopped in the second layer were identified by plotting
the energy deposited in the first layer (${\it \Delta}E$) against
the energy deposited in the second layer ($E$), as shown in 
Fig.~\ref{fig:2dadc}(a).
Protons were distinguished from heavier particles such as deuterons and
tritons using a particle-identification function defined by
\begin{equation}
f_{\rm PI}(E)=\frac{{\it \Delta}E-\varepsilon_{\rm p}(E)}
{\varepsilon_{\rm t}(E)-\varepsilon_{\rm p}(E)},
\end{equation}
where $\varepsilon_{\rm p}(E)$ and $\varepsilon_{\rm t}(E)$ 
are the calculated energy deposits at the first layer for a proton and a triton
having the energy deposit $E$ at the second layer.
The cut point of $f_{\rm PI}$ is shown as a curve 
in Fig.~\ref{fig:2dadc}(a).
Under this cut condition, typically 99\% of protons stopping in the 
second layer was accepted and a typical background rate due to
misidentified deuterons was 3\%, 
slightly depending on the detector angle and the kinematical conditions.
As seen in Fig.~\ref{fig:2dadc}(a), 
particles with a small energy deposit in the first and second layer were
clearly removed as background events due to electrons.
Background electrons which reached the third layer were excluded from
the plots due to their smaller energy deposit in the first and second layers.
Particles which reached the third layer are dominated by protons.
In this case, a proton was clearly identified by using
energies deposited in the second and third layer
(Fig.~\ref{fig:2dadc}(b)).

The kinetic energy of emitted charged particles from the target
was determined by summing the energy deposited
in the stack of SSD layers.
For particles which did not stop in the stack,
the kinetic energy was estimated from the $dE/dx$ 
measurements, assuming such particles to be protons.
The recoil energy $E_{\rm R}$ of the residual system was 
deduced from the measured 
momenta of the electron and proton.
Then the missing-energy $E_{\rm m}$ is calculated as
\begin{equation}
 E_{\rm m}=\omega-E_{\rm p}-E_{\rm R},
\end{equation}
where $E_{\rm p}$ is the kinetic energy of the emitted proton.
The final state of the reaction is defined by this $E_{\rm m}$.
A triple differential cross sections 
${d^{3}\sigma}/{d\omega d{\it \Omega}_{\rm e}d{\it \Omega}_{\rm p}}$
corresponding to specific final states were obtained by integrating proton
yield over a certain range of $E_{\rm m}$.
The normalization of the cross section was made from elastic
scattering by comparing the data with that of reference~\cite{li71}.

In general,
the (e,e$'$p) reaction cross section is decomposed into four 
terms~\cite{kle83};
\begin{equation}
 \frac{d^{3}\sigma}{d\omega d{\it \Omega}_{\rm e}d{\it \Omega}_{\rm p}}=
\sigma_{\rm L}+\sigma_{\rm T}+
\sigma_{\rm LT}\cos\phi_{\rm p}+\sigma_{\rm TT}\cos 2\phi_{\rm p},
\label{eex-xs}
\end{equation}
where $\sigma_{\rm L}$ and $\sigma_{\rm T}$ are longitudinal and transverse
terms, and the $\sigma_{\rm LT}$ and $\sigma_{\rm TT}$
are longitudinal-transverse and transverse-transverse
interference terms, respectively.
In the present experiment, protons were measured 
at $\phi_{\rm p}=-45^{\circ},-135^{\circ}$.
Therefore, the $\sigma_{\rm TT}\cos 2\phi_{\rm p}$ term is always zero.
The $\sigma_{\rm LT}\cos\phi_{\rm p}$ term contributes with opposite signs 
to the the cross sections
at $\phi_{\rm p}=-45^{\circ} $ and $ -135^{\circ}$, and 
does not contribute to a total 
cross section.
In the analysis, the data were fitted with  a linear combination of 
Legendre polynomials
\begin{equation}
\frac{d^{3}\sigma}{d\omega d{\it \Omega}_{\rm e}d{\it \Omega}_{\rm p}}=
A_{0} \left(1+ \sum_{i=1}^{m} a_{i} P_{i}
(\cos \theta_{\rm p})+\cos\phi_{\rm p} \sum_{j=1}^{n} b_{j} P^{1}_{j}
(\cos \theta_{\rm p}) \right).
\label{leg}
\end{equation}
The first and second terms in the equation~(\ref{leg}) correspond to 
$\sigma_{\rm L}+\sigma_{\rm T}$ terms, 
and the third term corresponds to the $\sigma_{\rm LT}\cos\phi_{\rm p}$ 
term~\cite{kle83}.
A least $\chi^{2}$ fitting was carried out under the constraint that
a triple differential cross section 
is not negative.
The maximum order of fitting polynomials $m$ and $n$ were terminated
when the $\chi^{2}$ per the number of degrees of freedom 
($\chi^{2}/NDF$) did not decrease with a larger value of $m$ or $n$.  

By integrating a triple differential cross section 
${d^{3}\sigma}/{d\omega d{\it \Omega}_{\rm e}d{\it \Omega}_{\rm p}}$
over proton emission angles, the double differential
cross section ${d^{2}\sigma}/{d\omega d{\it \Omega}_{\rm e}}$ 
was obtained as
\begin{equation}
 \frac{d^{2}\sigma}{d\omega d{\it \Omega}_{\rm e} }=4\pi A_{0}.
\label{4piA0}
\end{equation}

\section{Results and Discussions}

\begin{figure}
\begin{center}
\includegraphics{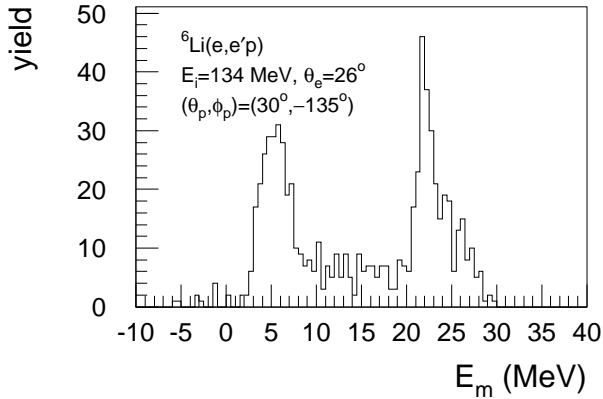}
\caption{Missing-energy spectrum for \nuc{6}{Li}(e,e$'$p) reaction.}
\label{fig:em-spectra}
\end{center}
\end{figure}

Figure~\ref{fig:em-spectra} shows an example of a missing-energy
spectrum.
Considering the level structure of \nuc{5}{He}~\cite{ajz88}, 
the peak at $E_{\rm m}=4.6$~MeV
corresponds to the 
two-body breakup into proton and \nuc{5}{He} (g.s., $J^{\pi}=3/2^{-}$).
A \nuc{5}{He} nuclei is an unstable nuclei, which has a ground state
as a resonance with ${\it \Gamma} = 600$~keV at 0.89~MeV above 
the $\alpha + {\rm n}$ threshold. 
The peak at $E_{\rm m}=21.4$~MeV corresponds to 
an excited state of \nuc{5}{He} with $J^{\pi}=3/2^{+}$
at 60~keV above the ${\rm d}+{\rm t}$
threshold with a width of 60~keV.
Events of three body
breakup reaction, \nuc{6}{Li}$\to \alpha+$n$+$p appear in the $E_{\rm m}$
region from 4.6 to 21.4~MeV. 
In the region of $E_{\rm m}\ge 21$~MeV, \nuc{6}{Li}$\to$ p$+$d$+$t
process contribute to the reaction. 
The reaction cross sections were obtained by integrating (e,e$'$p) yields
over the following three $E_{\rm m}$ regions.

Region-1: $2 < E_{\rm m} \le 8$ MeV, 

Region-2: $8 < E_{\rm m} \le 20$ MeV, 

Region-3: $20 < E_{\rm m} \le 23$ MeV. 

For each $E_{\rm m}$ region,
the measured triple differential cross sections for the \nuc{6}{Li}(e,e$'$p)
reaction are listed in Tables~\ref{tab:xs-1},
\ref{tab:xs-2} and \ref{tab:xs-3}.

\begin{figure}
\begin{center}
\includegraphics{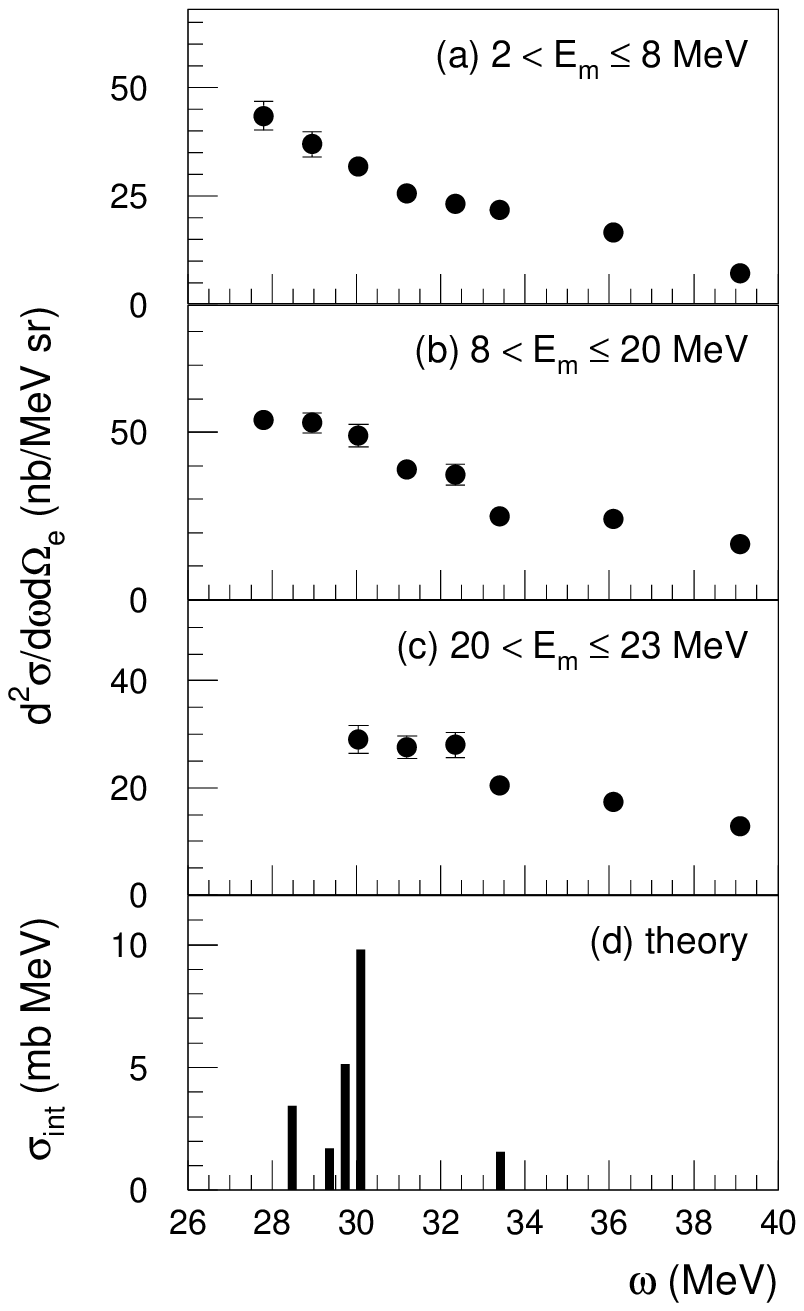}
\caption{The \nuc{6}{Li}(e,e$'$p) cross section 
at $\theta_{\rm e}=26^{\circ}$, integrated over proton
emission angle. (a) for $E_{\rm m}$ region 1, (b) for 2, and (c) for 3.
The calculated total photo absorption cross section from
Ref.~\cite{kur70} is shown in (d).}
\label{fig:totxs}
\end{center}
\end{figure}

By using Eq.~(\ref{4piA0}),
the double differential \nuc{6}{Li}(e,e$'$p) cross sections were
obtained from fitting of the triple differential cross sections 
with a function~(\ref{leg}). 
The results of the fitting are listed in Table~\ref{tab:fit}.
For every $E_{\rm m}$ region, 
the double differential \nuc{6}{Li}(e,e$'$p) cross sections 
smoothly decrease with $\omega$, as shown in Fig.~\ref{fig:totxs}.

In Fig.~\ref{fig:totxs}(c), data points for $\omega < 30$~MeV
do not exist since integration of the triple differential 
cross section over $\theta_{\rm p}$ was not possible because of
the detection threshold for protons.
However, cross sections at forward angles which are listed in 
the Table~\ref{tab:xs-3}, also show a smooth $\omega$ dependence for
$\omega < 30$~MeV.
No distinct structure of the GDR is observed.
Such smooth $\omega$ dependence is similar to the photoreaction 
cross section for $E_{\gamma}<30$~MeV measured with monochromatic 
photons~\cite{jun79,ber75,ahr75}, 
and contrasts with the (e,e$'$p) reactions for heavier 
nuclei~\cite{cal94,neu94} in the GDR energy regions,
where an $\omega$ dependence of the peak of the GDR was observed.
A theoretical prediction of the GDR states which decay into the
$3/2^{+}$ state of \nuc{5}{He} at $E_{\rm m}=21$~MeV is shown
in Fig.~\ref{fig:totxs}(d).

\begin{figure}
\begin{center}
\includegraphics{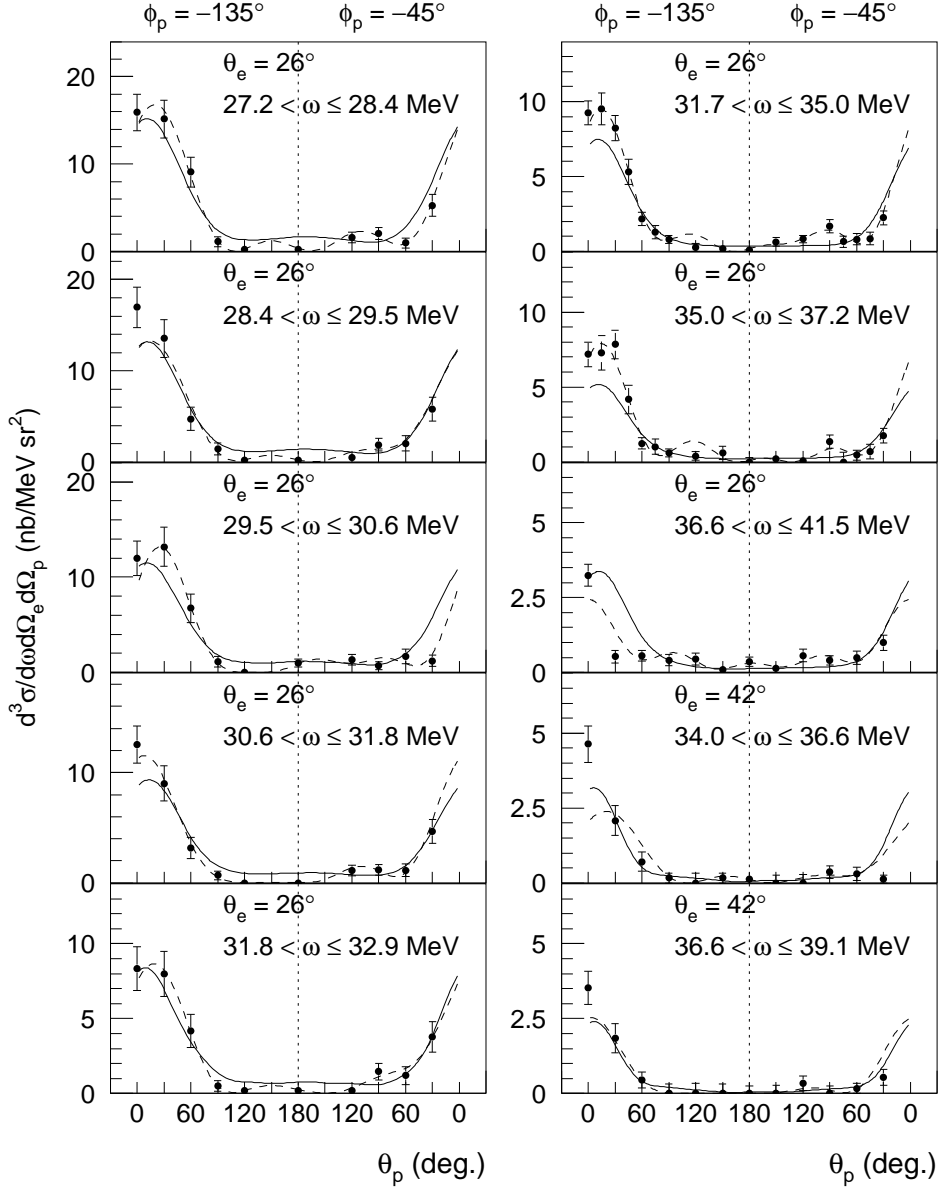}
\caption{Angular distributions of 
the triple differential cross sections for the 
\nuc{6}{Li}(e,e$'$p) reactions for $E_{\rm m}$ region-1 
($2<E_{\rm m}\le 8$~MeV). 
Solid curves show the result of the DWIA calculation multiplied by
0.50. The dashed curves show the fitting with the Legendre polynomials.}
\label{fig:dw-1}
\end{center}
 \end{figure}

\begin{figure}
\begin{center}
\includegraphics{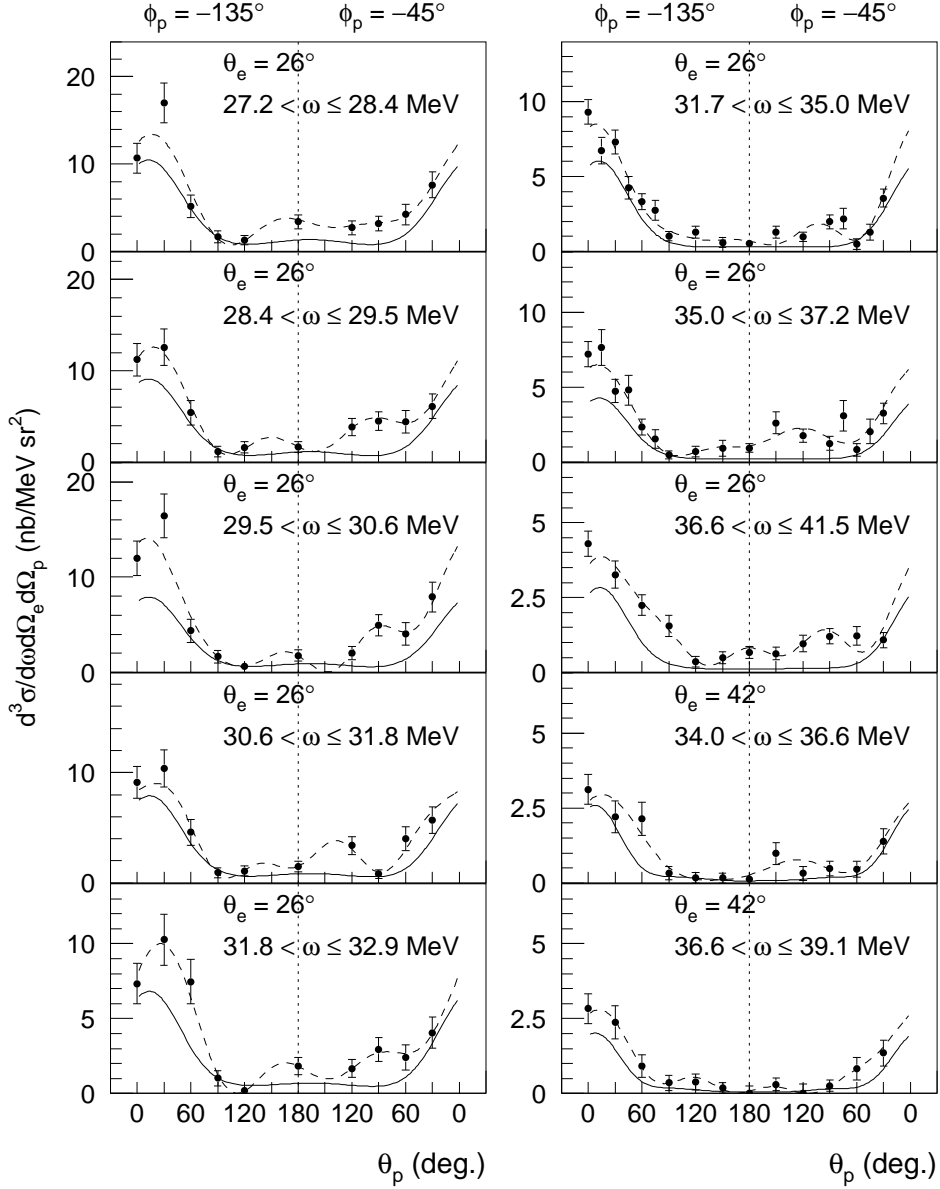}
\caption{Same as Fig.~\ref{fig:dw-1}, but for $E_{\rm m}$ region-2
($8<E_{\rm m}\le 20$~MeV).} 
\label{fig:dw-2}
\end{center}
\end{figure}

\begin{figure}
\begin{center}
\includegraphics{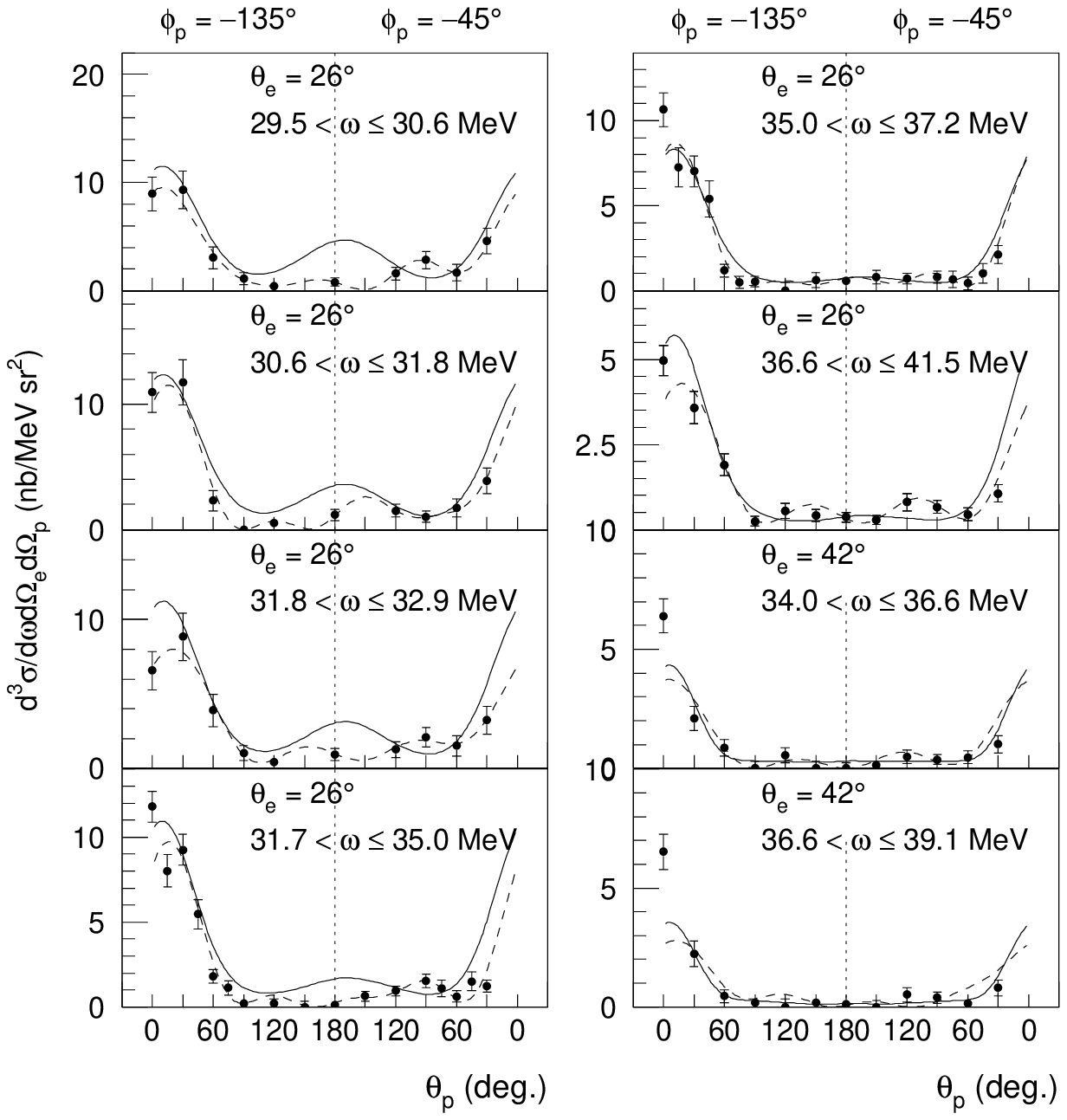}
\caption{Same as Fig.~\ref{fig:dw-1}, but for $E_{\rm m}$ region-3
($20<E_{\rm m}\le 23$~MeV).} 
\label{fig:dw-3}
\end{center}
\end{figure}

The angular distributions for each $E_{\rm m}$ regions are
shown in Figs.~\ref{fig:dw-1}, \ref{fig:dw-2}, and \ref{fig:dw-3}.
Results of the fitting with function~(\ref{leg}) are displayed
by dashed curves.
\input{table1}

\input{table2}
\input{table3}
\input{table4a}
\input{table4b}

For every final state, triple differential cross sections 
are large at small angles of $\theta_{\rm p}$, the proton angle with
respect to the momentum transfer.
This characteristic is observed over all the measured energy and momentum
region.
The angular distribution is different from that for heavier nuclei.
In the cases of \nuc{12}{C} and \nuc{40}{Ca}, 
protons are strongly emitted to $\theta_{\rm p}=0^{\circ}$ and $180^{\circ}$,
which indicates that the excitation of the dipole states dominate the
reaction mechanism.
The observed angular distribution with a peak at forward angles indicates
that the dipole excitation is not a dominant process, 
and higher multipoles contribute to the reaction.
In the present statistics and the number of data points,
most of data are well fitted within the fourth order
of Legendre polynomials.
However, for the $E_{\rm m}$ region-1 and 3,
cross sections at the forward angle excess the Legendre curves
at higher energies and momentum transfers.
Higher multipoles than the quadrupole may contribute to the reaction. 

The cross sections decrease smoothly with $\omega$ and the angular 
distributions show a forward peak.
It implies that the direct proton knockout
process dominates the \nuc{6}{Li}(e,e$'$p) reaction.
Therefore, we compared the measured triple differential cross section 
with a distorted wave impulse approximation (DWIA) calculation
which assumes a direct transition of a proton from the bound states 
to continuum states with distortion due to the 
interaction with the residual nucleus~\cite{dweepy}.
Ingredients of the calculation are then the bound state wave function 
of a proton in \nuc{6}{Li} and the optical potential between the emitted
proton and the residual nucleus. 
Although the \nuc{6}{Li} ground state wave function with a large
shell-model space based on a microscopic cluster model~\cite{lov90}
was used for a precise analysis of quasi-elastic
\nuc{6}{Li}(e,e$'$p) data~\cite{lan89},
we used a linear combination of harmonic
oscillator wave functions for 1s and 1p states
to simplify the calculation.
The strengths of these states were adjusted to
reproduce the \nuc{6}{Li}(e,e$'$p) cross sections in the quasi-elastic
region~\cite{lan89}.
Parameters of the wave functions used in the calculation are listed in
Table~\ref{tab:parm-dwia}.
\begin{table}
\caption{Parameters of harmonic oscillator wave functions used for
the calculation.}
\label{tab:parm-dwia}
\begin{tabular}{cccc}\hline
& \multicolumn{2}{c}{relative strength} &  
\\ \cline{2-3}
& 1s & 1p &  $b$ (fm) \\ \hline
$2<E_{\rm m}\le 8$ MeV & 0.3 & 0.25 & 2.03 \\
$8<E_{\rm m}\le 9.7$ MeV & 0.18 & 0.13 & 2.03 \\
$9.7<E_{\rm m}\le 20$ MeV & 0.19 & 0.0 & 1.5 \\
$20<E_{\rm m}\le 23$ MeV & 0.35 & 0.03 & 2.03 \\ \hline
\end{tabular}
\end{table}
The size parameter of the harmonic oscillator $b=2.03$~fm was taken 
from the value determined by elastic and inelastic electron scattering 
experiment on \nuc{6}{Li}~\cite{don73}.
Following the analysis of the quasi-elastic experiment~\cite{lan89},
the $E_{\rm m}$ region-2 was decomposed into two parts in the calculation
($8<E_{\rm m}\le 9.7$~MeV and $9.7<E_{\rm m}\le 20$~MeV).
In order to reproduce the quasi-elastic data for
$9.7 < E_{\rm m} \le 20$~MeV, the $b$ parameter was adjusted to 1.5~fm.
Energy-dependent optical potential parameters were used in the calculation,
which were determined from proton scattering 
in the energy range of 10--50 MeV for 
1p-shell nuclei~\cite{wat69}. 

The results of the DWIA calculation are compared with the experimental
cross section in Fig.~\ref{fig:dw-1}--\ref{fig:dw-3}.
For every $E_{\rm m}$ region, the cross section obtained from the DWIA
calculation exceeds the measured cross section.
For the comparison, the DWIA results were multiplied by a common 
normalization factor obtained from the least $\chi^{2}$ fitting 
of data for every $E_{\rm m}$ region and measured energy
and momentum transfer region.
A normalization factor 0.50 was found to give an overall agreement of the
DWIA result with the experimental data of both the angular distribution
and the $\omega$ dependence.
Comparing with measured data of $|\,\vec{q}\,|=90~{\rm MeV}/c$ at
$\theta_{\rm e}=42^{\circ}$ and 60--67~MeV/$c$ at 
$\theta_{\rm e}=26^{\circ}$, the DWIA calculation also reproduced
the $|\,\vec{q}\,|$-dependence of the cross section.
For $\theta_{\rm p}=30^{\circ}$ and $60^{\circ}$, larger cross 
sections were observed for $\phi_{\rm p}=-135^{\circ}$ than for
$\phi_{\rm p}=-45^{\circ}$ (Fig.~\ref{fig:dw-1}--\ref{fig:dw-3}).
As explained in the previous section, this difference is due to the
contribution of the longitudinal-transverse interference term,
$\sigma_{\rm LT}$. 
The DWIA calculation also reproduced the $\sigma_{\rm LT}$
contributions.

\section{Conclusion}
The \nuc{6}{Li}(e,e$'$p) reaction cross sections have been measured 
at ($\theta_{\rm e}=26^{\circ}, 27.2\le\omega\le 46.4$ MeV,
$|\,\vec{q}\,|$=60--67~MeV)
and ($\theta_{\rm e}=42^{\circ}, 34.0\le\omega\le 39.1$ MeV,
$|\,\vec{q}\,|$=90~MeV),
and were separated into three $E_{\rm m}$ regions
corresponding to different final states:
(1) residual \nuc{5}{He} in $3/2^{-}$ ground state,
(2) three-body breakup into $\alpha+{\rm n}+{\rm p}$,
and (3) residual \nuc{5}{He} in $3/2^{+}$, 16.7~MeV state.
Although dipole states which decay into $3/2^{+}$ state 
were predicted around 30~MeV~\cite{kur70},
the double differential cross section 
decreases smoothly with $\omega$, and no distinct peak of the expected GDR
is observed for every final state. 
This smooth $\omega$-dependence is similar to 
photoreaction results for $E_{\gamma}<30$ MeV except that 
of Denisov {\it et al.}~\cite{den67}.
Angular distributions of triple differential cross sections have 
a peak at the direction of momentum transfer.
This common characteristic is observed 
in the entire region of the measured energy and momentum transfer, 
in contrast with the results of other (e,e$'$p) reactions on
$^{12}$C and $^{40}$Ca where the GDR is clearly 
observed.
On the other hand, the DWIA calculation
reproduces well the experimental data with only one normalization factor.
This agreement indicates that the direct process dominates
the \nuc{6}{Li}(e,e$'$p) reactions in this energy and momentum 
transfer region.

In order to study the (e,e$'$p) reaction mechanism more precisely,
further calculations with a microscopic cluster 
model~\cite{lov90} for the \nuc{6}{Li} ground state wave function
might be needed.

\begin{ack}
We express our thanks to the accelerator crew and staffs of the
Laboratory of Nuclear Science, Tohoku University for their excellent 
operation of the accelerator and the experimental apparatus.

We would like to thank Prof. C. Giusti and Prof. G. van der
Steenhoven for helping us to carry out the DWIA calculation 
with the computer code DWEEPY.

We would also like to thank Prof. G.A. Peterson for a careful 
reading of the manuscript.
\end{ack}

\end{document}

%% file: table1.tex
\landscape
\begin{table}
\caption{The triple differential \nuc{6}{Li}(e,e$'$p) cross section 
for $E_{\rm m}$ region-1 ($2<E_{\rm m}\le 8$~MeV), 
in units of nb/MeV sr$^{2}$.
Cross sections displayed with ``$<$'' sign are Poisson upper limits at a 
68\% confidence level, where no event was observed.}
\label{tab:xs-1}
\begin{tabular}{cccccccccccc}\hline
& \multicolumn{11}{c}{$\theta_{\rm e}$ and $\omega$ (MeV)} \\ \cline{2-12}
& $26^{\circ}$ & $26^{\circ}$ & $26^{\circ}$ & $26^{\circ}$ & $26^{\circ}$ &
$26^{\circ}$ & $26^{\circ}$ & $26^{\circ}$ & $26^{\circ}$& $42^{\circ}$ & 
$42^{\circ}$ \\ 
$(\theta_{\rm p},\phi_{\rm p})$&  27.2--28.4 & 28.4--29.5 &
29.5--30.6 & 30.6--31.8 & 31.8--32.9 & 31.7--35.0 & 35.0--37.2 & 
36.6--41.5 & 41.7--46.6 & 34.0--36.6 & 36.6--39.1 \\ \hline
$(0^{\circ},-)$ & $15.9\pm 2.1$ & $17.0\pm 2.2$ & $12.0\pm 1.8$ &
$12.5\pm 1.7$ & $8.3\pm 1.5$ & 
$9.2\pm 0.8$ & $7.2\pm 0.8$ &
$3.2\pm 0.4$ & $1.5\pm 0.3$ & $4.6\pm 0.6$ & $3.5\pm 0.5$ \\
$(15^{\circ},-135^{\circ})$ & $-$ & $-$ & $-$ & $-$ & $-$ &
$9.5\pm 1.1$ & $7.3\pm 1.2$ & $-$ & $-$ &  $-$ & $-$ \\
$(30^{\circ},-135^{\circ})$ & $15.2\pm 2.1$ & $13.5\pm 2.1$ &
$13.2 \pm 2.1$ & $9.0\pm 1.6$ & $8.0\pm 1.5$  &$8.2\pm 0.8$ &
$7.8\pm 1.0$ & $0.5\pm 0.2$ & $1.0\pm 0.3$ & $2.1\pm 0.5$ & $1.8\pm 0.5$ \\
$(45^{\circ},-135^{\circ})$ & $-$ & $-$ & $-$ & $-$ & $-$ &
$5.3\pm 0.8$ & $4.2\pm 0.9$ &  $-$ & $-$ & $-$ & $-$ \\
$(60^{\circ},-135^{\circ})$ & $9.1\pm 1.7$ & $4.7\pm 1.3$ &
$6.7\pm 1.5$ & $3.2\pm 1.0$ & $4.2\pm 1.1$ & $2.2\pm 0.4$ &
$1.2\pm 0.4$ & $0.6\pm 0.2$ & $0.6\pm 0.2$ & $0.7\pm 0.3$ & $0.5\pm 0.3$ \\
$(75^{\circ},-135^{\circ})$ &  $-$ & $-$ & $-$ & $-$ & $-$ &
$1.3\pm 0.5$ & $1.0\pm 0.5$ & $-$ & $-$ & $-$ & $-$ \\
$(90^{\circ},-135^{\circ})$ & $1.1\pm 0.6$ & $1.5\pm 0.7$ & $1.1\pm 0.6$ &
$0.7\pm 0.4$ & $0.5\pm 0.4$ & 
$0.8\pm 0.3$ & $0.6\pm 0.3$ &
$0.4\pm 0.2$ & $-$ &  $0.2\pm 0.2$ & $<0.32$ \\
$(120^{\circ},-135^{\circ})$ & $0.2\pm 0.2$ & $0.2\pm 0.2$ &
$<0.39$ & $<0.33$ & $0.2\pm 0.2$ &
$0.3 \pm 0.3$ & $0.4\pm 0.3$ &
$0.4\pm 0.2$ & $-$ & $<0.33$ & $<0.35$ \\
$(150^{\circ},-135^{\circ})$&  $-$ & $-$ & $-$ & $-$ & $-$ &
$0.2\pm 0.2$ & $0.6 \pm 0.4 $ & $0.1\pm 0.1$ & $-$ & $0.2\pm 0.2$ &
$<0.32$ \\
$(180^{\circ},-)$ & $0.2\pm 0.2$ & $0.2\pm 0.2$ & $1.0\pm 0.4$ &
$<0.30$ & $0.18\pm 0.18$ &
$0.1\pm 0.1$ & $<0.31$ & $0.4\pm 0.1$ &
$-$ & $0.1\pm 0.1$ & $<0.24$ \\
$(30^{\circ},-45^{\circ})$ & $5.3\pm 1.2$ & $5.8\pm 1.3$ &
$1.2\pm 0.6$ & $4.7\pm 1.1$ & $3.8\pm 1.0$ & 
$2.2\pm 0.5$ &
$1.8\pm 0.5$ & $1.0\pm 0.3$ & $0.1\pm 0.1$ & $0.1\pm 0.1$ & $0.5\pm 0.3$ \\
$(45^{\circ},-45^{\circ})$ &  $-$ & $-$ & $-$ & $-$ & $-$ &
$0.9\pm 0.4$ & $0.7\pm 0.5$ & $-$ & $-$ &  $-$ & $-$ \\
$(60^{\circ},-45^{\circ})$ & $1.0\pm 0.6$ & $2.0\pm 0.8$ &
$1.7\pm 0.8$ & $1.2\pm 0.6$ & $1.2\pm 0.6$ &
$0.8\pm 0.4$ &
$0.5\pm 0.3$ & $0.5\pm 0.2$ & $0.2\pm 0.1$ & $0.3\pm 0.2$ & $0.2\pm 0.2$ \\
$(75^{\circ},-45^{\circ})$ & $-$ & $-$ & $-$ & $-$ & $-$ &
$0.7\pm 0.4$ & $<0.63$ & $-$ & $-$ &  $-$ & $-$ \\
$(90^{\circ},-45^{\circ})$ & $2.0\pm 0.7$ & $1.9\pm 0.7$ & $0.7\pm 0.4$ &
$1.2\pm 0.5$ & $1.5\pm 0.6$  &
$1.7\pm 0.4$ & $1.4\pm 0.4$ &
$0.4\pm 0.1$ & $0.1\pm 0.1$ & $0.4\pm 0.2$ & $<0.24$ \\
$(120^{\circ},-45^{\circ})$  &  $1.6\pm 0.6$ & $0.5\pm 0.3$ & $1.3\pm 0.6$ &
$1.1\pm 0.5$ & $0.2\pm 0.2$ &
$0.8\pm 0.3$ & $0.1\pm 0.1$ &
$0.6\pm 0.2$ & $-$ & $<0.29$ & $0.3\pm 0.2$ \\
$(150^{\circ},-45^{\circ})$ & $-$ & $-$ & $-$ & $-$ & $-$ & $0.6\pm 0.3$ &
$0.2\pm 0.2$ & $0.1\pm 0.1$ & $-$ & $<0.26$ & $<0.28$ \\ \hline
\end{tabular}
\end{table}
\endlandscape

%% file: table2.tex
\landscape
\begin{table}
\caption{Same as for Table~\ref{tab:xs-1}, 
but for $E_{\rm m}$ region-2 ($8<E_{\rm m}\le 20$~MeV).}
\label{tab:xs-2}
\begin{tabular}{cccccccccccc}\hline
& \multicolumn{11}{c}{$\theta_{\rm e}$ and $\omega$ (MeV)} \\ \cline{2-12}
& $26^{\circ}$ & $26^{\circ}$ & $26^{\circ}$ & $26^{\circ}$ & $26^{\circ}$ &
$26^{\circ}$ & $26^{\circ}$ & $26^{\circ}$ & $26^{\circ}$& $42^{\circ}$ & 
$42^{\circ}$ \\ 
$(\theta_{\rm p},\phi_{\rm p})$&  27.2--28.4 & 28.4--29.5 &
29.5--30.6 & 30.6--31.8 & 31.8--32.9 & 31.7--35.0 & 35.0--37.2 & 
36.6--41.5 & 41.7--46.6 & 34.0--36.6 & 36.6--39.1 \\ \hline
$(0^{\circ},-)$ & $10.7\pm 1.7$ & $11.2\pm 1.8$ & $12.0\pm 1.8$ &
$9.1\pm 1.4$ & $7.3\pm 1.4$ & 
$9.3\pm 0.8$ & $7.2\pm 0.8$ & 
$4.3\pm 0.4$ & $2.6\pm 0.4$ & $3.1\pm 0.5$ & $2.8\pm 0.5$ \\ 
$(15^{\circ},-135^{\circ})$ & $-$ & $-$ & $-$ & $-$ & $-$ &
$6.7\pm 0.9$ & $7.6\pm 1.2$ & $-$ &
$-$ & $-$ & $-$ \\
$(30^{\circ},-135^{\circ})$ & $17.0\pm 2.3$ & $12.6\pm 2.0$ & $16.4\pm 2.3$ & 
$10.4\pm 1.7$ & $10.3\pm 1.7$ &
$7.3\pm 0.8$ & $4.8\pm 0.8$ & 
$3.3\pm 0.4$ & $4.3 \pm 0.6 $&$2.2\pm 0.5$  & $2.4\pm 0.6$    \\
$(45^{\circ},-135^{\circ})$ & $-$ & $-$ & $-$ & $-$ & $-$ &
$4.3\pm 0.8$ & $4.8\pm 1.0$ 
& $-$ &  $-$ & $-$ & $-$ \\
$(60^{\circ},-135^{\circ})$ & $5.2\pm 1.3$ & $5.4\pm 1.4$ & 
$4.4\pm 1.2$ & $4.6\pm 1.1$ & $7.5\pm 1.5$ &
$3.3\pm 0.5$ & 
$2.3\pm 0.5$ & $2.2\pm 0.3$ & $1.4\pm 0.3$ & $2.1\pm 0.6$ & $0.9\pm 0.4$ \\
$(75^{\circ},-135^{\circ})$ & $-$ & $-$ & $-$ & $-$ & $-$ &
$2.7\pm 0.7$ & $1.5\pm 0.6$ & 
$-$ & $-$ & $-$ & $-$ \\
$(90^{\circ},-135^{\circ})$ & $1.7\pm 0.7$ & $1.2\pm 0.6$ & $1.7\pm 0.7$ & 
$0.9\pm 0.5$ & $1.0\pm 0.5$ &
$1.0\pm 0.3$ & $0.5\pm 0.3$ &
$1.6\pm 0.4$ & $-$ & $0.3\pm 0.2$ & $0.4\pm 0.2$ \\
$(120^{\circ},-135^{\circ})$ & $1.3\pm 0.5$ & $1.6\pm 0.6$ & 
$0.6\pm 0.4$ & $1.1\pm 0.4$ & $0.2\pm 0.2$ &
$1.3 \pm 0.4$ & $0.7\pm 0.3$ &
$0.4\pm 0.2$ & $-$ & $0.2\pm 0.2$ & $0.4\pm 0.3$ \\
$(150^{\circ},-135^{\circ})$ & $-$ & $-$ & $-$ & $-$ & $-$ &
$0.6\pm 0.3$ & $0.9 \pm 0.5 $ & 
$0.5\pm 0.2$ & $-$ & $0.2\pm 0.2$ & $0.2\pm 0.2$ \\
$(180^{\circ},-)$ & $3.4\pm 0.8$ & $1.7\pm 0.6$ & $1.8\pm 0.6$ &
$1.5\pm 0.5$ & $1.8\pm 0.6$ &
$0.5\pm 0.2$ & $0.9\pm 0.3$ & $0.7\pm 0.2$ &
$-$ & $0.1\pm 0.1$ & $<0.24$ \\
$(30^{\circ},-45^{\circ})$ & $7.6\pm 1.5$ & $6.1\pm 1.4$ & 
$7.9\pm 1.6$ & $5.7\pm 1.2$ & $4.1\pm 1.0$ &
$3.6\pm 0.6$ & 
$3.3\pm 0.7$ & $1.1\pm 0.3$ & $1.0\pm 0.3$ & $1.4\pm 0.4$ & $1.3\pm 0.4$ \\
$(45^{\circ},-45^{\circ})$ & $-$ & $-$ & $-$ & $-$ & $-$ &
$1.3\pm 0.5$ & $2.0\pm 0.8$ & 
$-$ & $-$ & $-$ & $-$ \\
$(60^{\circ},-45^{\circ})$ & $4.2\pm 1.2$ & $4.4\pm 1.2$ & 
$4.1\pm 1.2$ & $4.0\pm 1.1$ & $2.4\pm 0.9$ &
$0.5\pm 0.3$ &
$0.8\pm 0.4$ & $1.2\pm 0.3$ & $0.9\pm 0.3$ & $0.5\pm 0.3$ & $0.8\pm 0.4$ \\
$(75^{\circ},-45^{\circ})$ & $-$ & $-$ & $-$ & $-$ & $-$ &
$2.2\pm 0.7$ & $3.1\pm 1.0$ & 
$-$ & $-$ & $-$ & $-$ \\
$(90^{\circ},-45^{\circ})$ & $3.2\pm 0.8$ & $4.5\pm 1.0$ & $5.0\pm 1.1$ & 
$0.8\pm 0.4$ & $2.9\pm 0.8$ &
$2.0\pm 0.5$ & $1.2\pm 0.5$ & 
$1.2\pm 0.3$ & $0.2\pm 0.1$ & $0.5\pm 0.2$ & $0.3\pm 0.2$ \\
$(120^{\circ},-45^{\circ})$ &  $2.7\pm 0.8$ & $3.8\pm 1.0$ & $2.0\pm 0.7$ & 
$3.4\pm 0.8$ & $1.7\pm 0.6$ &
$1.0\pm 0.3$ & $1.8\pm 0.4$ &
$1.0\pm 0.3$ & $-$ & $0.3\pm 0.2$ & $<0.31$ \\
$(150^{\circ},-45^{\circ})$ & $-$ & $-$ & $-$ & $-$ & $-$ &
$1.3\pm 0.4$ &
$2.6\pm 0.7$ & $0.6\pm 0.2$ & $-$ & $1.0\pm 0.4$ & $0.3\pm 0.2$ \\ \hline
\end{tabular}
\end{table}
\endlandscape

%% file: table3.tex
\landscape
\begin{table}
\caption{Same as for Table~\ref{tab:xs-1}, 
but for $E_{\rm m}$ region-3 ($20<E_{\rm m}\le 23$~MeV).}
\label{tab:xs-3}
\begin{tabular}{cccccccccccc}\hline
& \multicolumn{11}{c}{$\theta_{\rm e}$ and $\omega$ (MeV)} \\ \cline{2-12}
& $26^{\circ}$ & $26^{\circ}$ & $26^{\circ}$ & $26^{\circ}$ & $26^{\circ}$ &
$26^{\circ}$ & $26^{\circ}$ & $26^{\circ}$ & $26^{\circ}$& $42^{\circ}$ & 
$42^{\circ}$ \\ 
$(\theta_{\rm p},\phi_{\rm p})$&  27.2--28.4 & 28.4--29.5 &
29.5--30.6 & 30.6--31.8 & 31.8--32.9 & 31.7--35.0 & 35.0--37.2 & 
36.6--41.5 & 41.7--46.6 & 34.0--36.6 & 36.6--39.1 \\ \hline
$(0^{\circ},-)$ & $8.8\pm 1.6$ & $9.5\pm 1.7$ & $9.0\pm 1.6$ & 
$10.9\pm 1.6$ & $6.6\pm 1.3$ &
$11.8\pm 0.9$ & $10.7\pm 1.0$ &$5.0\pm 0.5$ & $3.0\pm 0.4$ & $6.4\pm 0.7$ & 
$6.5\pm 0.7$\\
$(15^{\circ},-135^{\circ})$ & $-$ & $-$ & $-$ & $-$ & $-$ &
$8.0\pm 1.0$ & $7.3\pm 1.2$ & $-$ & $-$ &$-$ & $-$ \\
$(30^{\circ},-135^{\circ})$ & $9.3\pm 1.7 $ & $9.3\pm 1.7 $ & 
$9.3\pm 1.7$ & $11.8\pm 1.8$ & $8.8\pm 1.6$ & 
$9.3\pm 0.9$ & $7.0\pm 0.9$ &$3.6\pm 0.5$ & $2.2\pm 0.4$ & $2.1\pm 0.5$ & 
$2.2\pm 0.5$\\
$(45^{\circ},-135^{\circ})$ &  $-$ & $-$ & $-$ & $-$ & $-$ & 
$5.5\pm 0.9$ & $5.4\pm 1.1$ &  $-$ & $-$ & $-$ & $-$ \\
$(60^{\circ},-135^{\circ})$ & $-$ & $3.0\pm 1.0$ &  $3.0\pm 1.0$ 
& $2.3\pm 0.8$ & $3.9\pm 1.1$ &
$1.8\pm 0.4$ & $1.2\pm 0.4$ & $1.9\pm 0.3$ & 
$1.1\pm  0.3$&$0.9\pm 0.3$ & $0.5\pm 0.3$ \\
$(75^{\circ},-135^{\circ})$ & $-$ & $-$ & $-$ & $-$ & $-$ &
$1.1\pm 0.4$ & $0.5\pm 0.4$ & $-$ & $-$ & $-$ & $-$ \\
$(90^{\circ},-135^{\circ})$ & $-$ & $0.6\pm 0.4$ & $1.1\pm 0.6$ &
$<0.42$ & $1.0\pm 0.5$  &
$0.2\pm 0.2$ & $0.5\pm 0.3$ & $0.2\pm 0.1$ & $-$ &  $<0.30$ &
$0.2\pm 0.2$ \\
$(120^{\circ},-135^{\circ})$ & $-$ & $-$ & $0.4\pm 0.3$ & $0.5\pm 0.3$ & 
$0.4\pm 0.3$ & 
$0.2\pm 0.3$ & $<0.51$ & $0.5 \pm 0.2$ & $-$ & $0.5\pm 0.3$ &
$<0.35$ \\
$(150^{\circ},-135^{\circ})$ &  $-$ & $-$ & $-$ & $-$ & $-$ &
$<0.36$ & $0.6 \pm 0.4 $ & $0.4\pm 0.2$ &$-$ &  $<0.30$ & 
$0.2\pm 0.2$ \\
$(180^{\circ},-)$ & $-$ & $-$ & $0.8\pm 0.4$ & $1.2\pm 0.4$ & $0.9\pm 0.4$ & 
$0.1\pm 0.1$ & $0.6\pm 0.2$ & $0.4\pm 0.1$ &$-$ & $<0.22$ & $0.1\pm 0.1$\\
$(30^{\circ},-45^{\circ})$ & $2.3\pm 0.8$ & $1.8\pm 0.7$ & 
$4.6\pm 1.2$ & $3.9\pm 1.0$ & $3.2\pm 0.9$ &
$1.2\pm 0.4$ & $2.1\pm 0.5$ & $1.1\pm 0.3$ & 
$0.4\pm 0.2$&$1.0\pm 0.4$ & $0.8\pm 0.3$ \\
$(45^{\circ},-45^{\circ})$ & $-$ & $-$ & $-$ & $-$ & $-$ &
$1.5\pm 0.6$ & $1.0\pm 0.6$ & $-$ & $-$ & $-$ & $-$ \\
$(60^{\circ},-45^{\circ})$ & $-$ & $0.7\pm 0.5$ & $1.7\pm 0.8$ &
$1.7\pm 0.7$ & $1.5\pm 0.7$ &
$0.6\pm 0.3$ & $0.4\pm 0.4$ & $0.4\pm 0.2$ &
$0.7\pm 0.3$ & $0.5\pm 0.3$ & $0.2\pm 0.2$ \\
$(75^{\circ},-45^{\circ})$ & $-$ & $-$ & $-$ & $-$ & $-$ &
$1.1\pm 0.5$ & $0.7\pm 0.5$ & $-$ & $-$ & $-$ & $-$ \\
$(90^{\circ},-45^{\circ})$ & $-$ & $1.4\pm 0.6$&$2.8\pm 0.8$ & 
$1.0\pm 0.4$ & $2.1\pm 0.7$ & 
$1.5\pm 0.4$ & $0.8\pm 0.4$ & $0.7\pm 0.2$ & $0.2\pm 0.2$ & $0.4\pm 0.2$ & 
$0.4\pm 0.2$ \\
$(120^{\circ},-45^{\circ})$ &  $-$ & $-$ & $1.6\pm 0.6$ & 
$1.5\pm 0.5$ & $1.3\pm 0.5$  & 
$1.0\pm 0.3$ & $0.7\pm 0.3$ & $0.8\pm 0.3$ & $-$ & $0.5\pm 0.3$ &
$0.5\pm 0.3$ \\ 
$(150^{\circ},-45^{\circ})$ & $-$ & $-$ & $-$ & $-$ & $-$ &
$0.6\pm 0.3$ &
$0.8\pm 0.4$ & $0.3\pm 0.1$ & $-$ & $0.1\pm 0.1$ & $<0.28$ \\ \hline
\end{tabular}
\end{table}
\endlandscape

%% file: table4a.tex
\landscape
\begin{table}
\caption{Fitting parameters in the function (\ref{leg}).}
\label{tab:fit}
\tabcolsep=3.3pt
\hspace*{-1.5cm}
\begin{tabular}{cccccccccccc}\hline
$\theta_{\rm e}$ & $\omega$ (MeV) & $A_0$ (nb MeV$^{-1}$sr$^{-1}$) & $a_1$ & $a_2$ & $a_3$ & $a_4$ & $b_1$ & $b_2$ & $b_3$ & $b_4$ & $\chi^{2}/NDF$ \\ \hline
\multicolumn{11}{c}{Region 1; $2<E_{\rm m}\le 8$~MeV} \\
$26^{\circ}$ & 27.2-28.4 & $3.5\pm 0.3$ & $1.5\pm 0.1$ & $1.1\pm 0.2$ & $0.5\pm 0.1$ & $-$          & $0.4\pm 0.1$ & $0.53\pm 0.07$ & $0.30\pm 0.04$ & $-$ & 0.84\\
$26^{\circ}$ & 28.4-29.5 & $2.9\pm 0.2$ & $1.6\pm 0.1$ & $1.1\pm 0.1$ & $0.4\pm 0.1$ & $-$          & $0.2\pm 0.1$ & $0.31\pm 0.07$ & $0.18\pm 0.04$ & $-$ & 3.37\\
$26^{\circ}$ & 29.5-30.6 & $2.5\pm 0.2$ & $1.5\pm 0.2$ & $1.0\pm 0.2$ & $0.1\pm 0.1$ & $-$         & $0.45\pm 0.07$ & $0.68\pm 0.07$ & $0.36\pm 0.07$ & $0.14 \pm 0.04 $ & 3.98\\
$26^{\circ}$ & 30.6-31.8 & $2.0\pm 0.2$ & $1.7\pm 0.1$ & $1.4\pm 0.1$ & $1.1\pm 0.1$ & $0.32 \pm 0.09$ & $0.12\pm 0.08$ & $0.38\pm 0.07$ & $0.17\pm 0.03$ & $-$ & 1.46\\
$26^{\circ}$ & 31.8-32.9 & $1.8\pm 0.2$ & $1.7\pm 0.2$ & $1.1\pm 0.2$ & $0.3\pm 0.1$ & $-$          & $0.2\pm 0.1$ & $0.32\pm 0.09$ & $0.27\pm 0.08$ & $0.05\pm 0.04$ & 0.95 \\
$26^{\circ}$ & 31.7-35.0 & $1.73\pm 0.08$ & $1.30\pm 0.06$ & $1.05\pm 0.09$ & $1.10\pm 0.09$ & $0.44\pm 0.09$ & $0.26\pm 0.05$ & $0.35\pm 0.04$ & $0.30\pm 0.02$ & $0.16\pm0.02$ & 1.23\\
$26^{\circ}$ & 35.0-37.2 & $1.31\pm 0.08$ & $1.37\pm 0.08$ & $1.2\pm 0.1$ & $1.2\pm 0.1$ & $0.5\pm 0.1$ & $0.42\pm 0.07$ & $0.24\pm 0.05$ & $0.35\pm 0.04$ & $0.23\pm 0.03$ & 2.31\\
$26^{\circ}$ & 36.6-41.5 & $0.57\pm 0.05$ & $0.9\pm 0.1$ & $0.6\pm 0.2$ & $1.0\pm 0.2$ & $0.9\pm 0.2$  & $0.05\pm 0.1$ & $0.03\pm 0.06$ & $-0.03\pm 0.06$ & $-$ & 5.02\\
$42^{\circ}$ & 34.0-36.6 & $0.52\pm 0.07$ & $1.6\pm 0.3$ & $1.1\pm 0.3$ & $0.3\pm 0.2$ & $-$          & $0.4\pm 0.1$ & $0.4\pm 0.1$ & $0.27\pm 0.08$ & $-$ & 6.56\\
$42^{\circ}$ & 36.6-39.1 & $0.34\pm 0.05$ & $2.0\pm 0.4$ & $2.1\pm 0.3$ & $1.6\pm 0.4$ & $0.6\pm 0.2$ & $0.1\pm 0.1$ & $0.3\pm 0.1$ & $0.13\pm 0.07$ & $-$ & 2.86\\ \hline
\multicolumn{11}{c}{Region 2; $8<E_{\rm m}\le 20$~MeV} \\
$26^{\circ}$ & 27.2-28.4 & $4.3\pm 0.2$ & $0.9\pm 0.1$ & $0.9\pm 0.1$ & $0.1\pm 0.1$ & $-$          & $0.02\pm 0.08$ & $0.20\pm 0.07$ & $0.15\pm 0.06$ & $-$ & 2.23\\
$26^{\circ}$ & 28.4-29.5 & $4.2\pm 0.2$ & $0.9\pm 0.1$ & $0.5\pm 0.1$ & $0.3\pm 0.1$ & $-$          & $-0.10\pm 0.08$ & $0.19\pm 0.06$ & $0.22\pm 0.07$ & $-$ & 0.24\\
$26^{\circ}$ & 29.5-30.6 & $3.9\pm 0.3$ & $1.2\pm 0.1$ & $0.6\pm 0.1$ & $0.3\pm 0.1$ & $0.4\pm 0.2$ & $-0.07\pm 0.08$ & $0.15\pm 0.05$ & $0.22\pm 0.05$ & $-$ & 2.83 \\ 
$26^{\circ}$ & 30.6-31.8 & $3.1\pm 0.2$ & $0.9\pm 0.2$ & $1.1\pm 0.1$ & $0.2\pm 0.2$ & $-0.5\pm 0.2$ & $0.01\pm 0.08$ & $0.25\pm 0.07$ & $0.02\pm 0.06$ & $-$ & 1.53\\ \hline
\end{tabular}
\end{table}
\endlandscape

%% file: table4b.tex
\landscape
\begin{table}
\contcaption{}
\tabcolsep=3.3pt
\hspace*{-1.5cm}
\begin{tabular}{cccccccccccc}\hline
$\theta_{\rm e}$ & $\omega$ (MeV) & $A_0$ (nb MeV$^{-1}$sr$^{-1}$) & $a_1$ & $a_2$ & $a_3$ & $a_4$ & $b_1$ & $b_2$ & $b_3$ & $b_4$ & $\chi^{2}/NDF$ \\ \hline
\multicolumn{11}{c}{Region 2; $8<E_{\rm m}\le 20$~MeV -- continued.} \\
$26^{\circ}$ & 31.8-32.9 & $3.0\pm 0.3$ & $1.1\pm 0.1$ & $0.6\pm 0.2$ & $-0.1\pm 0.1$ & $-$ & $0.1\pm 0.1 $ & $0.33\pm 0.08$ & $0.24\pm 0.07$ & $-$ & 0.40\\
$26^{\circ}$ & 31.7-35.0 & $2.0\pm 0.1$ & $1.09\pm 0.08$ & $0.8\pm 0.1$ & $0.8\pm 0.1$ & $0.4\pm 0.1$ & $-0.17\pm 0.07 $ & $0.27\pm 0.05$ & $0.16\pm 0.04$ & $-$ & 2.42\\
$26^{\circ}$ & 35.0-37.2 & $1.9\pm 0.1$ & $0.8\pm 0.1$ & $0.9\pm 0.1$ & $0.6\pm 0.1$ & $-$ & $-0.12\pm 0.07$ & $0.27\pm 0.06$ & $0.10\pm 0.05$ & $-$ & 1.55\\
$26^{\circ}$ & 36.6-41.5 & $1.32\pm 0.07$ & $0.82\pm 0.08$ & $0.2\pm 0.1$ & $0.2\pm 0.1$ &$0.4\pm0.1$ & $0.19\pm 0.06 $ & $0.32\pm 0.04$ & $0.12\pm 0.03$ & $-$ & 2.82\\
$42^{\circ}$ & 34.0-36.6 & $0.82\pm 0.09$ & $1.1\pm 0.2$ & $0.8\pm 0.2$ & $0.4\pm 0.2$ & $-$           & $0.2\pm 0.1 $ & $0.4\pm 0.1$ & $0.08\pm 0.07$ & $-$ & 1.54\\
$42^{\circ}$ & 36.6-39.1 & $0.65\pm 0.06$ & $1.4\pm 0.2$ & $1.1\pm 0.2$ & $0.6\pm 0.2$ & $-$           & $0.2\pm 0.1 $ & $0.0\pm 0.1$ & $0.09\pm 0.07$ & $0.1\pm 0.06$ & 0.53\\ \hline
\multicolumn{11}{c}{Region 3; $20<E_{\rm m}\le 23$~MeV} \\
$26^{\circ}$ & 29.5-30.6 & $2.3\pm 0.2$ & $1.2\pm 0.1$ & $0.7\pm 0.2$ & $0.6\pm 0.2$ & $0.4\pm 0.2$ & $-0.01\pm 0.08 $& $0.25\pm 0.08$ & $0.24\pm 0.06$ & $-$ & 0.42 \\
$26^{\circ}$ & 30.6-31.8 & $2.2\pm 0.2$ & $1.3\pm 0.1$ & $1.6\pm 0.1$ & $0.8\pm 0.1$ & $-$          & $0.06\pm 0.07$ & $0.42\pm 0.07$ & $0.19\pm 0.05$ & $0.18\pm 0.03$ & 1.28\\
$26^{\circ}$ & 31.8-32.9 & $2.2\pm 0.2$ & $1.1\pm 0.2$ & $0.8\pm 0.2$ & $0.2\pm 0.2$ & $-$ & $0.2\pm 0.1 $ & $0.31\pm 0.09$ & $0.3\pm 0.1$ & $-$ & 0.46\\
$26^{\circ}$ & 31.7-35.0 & $1.62\pm 0.09$ & $1.41\pm 0.08$ & $1.23\pm 0.08$ & $1.06\pm 0.08$ & $0.41\pm 0.05$ & $0.20\pm 0.04$  & $0.47\pm 0.04$ & $0.40\pm 0.03$ & $0.19\pm 0.02$ & 5.06\\
$26^{\circ}$ & 35.0-37.2 & $1.4\pm 0.1$ & $1.4\pm 0.1$ & $1.5\pm 0.1$ & $1.2\pm 0.2$ & $0.7\pm 0.2$ & $0.21\pm 0.09 $ & $0.37\pm 0.06$ & $0.30\pm 0.05$ & $0.14\pm 0.05$ & 2.46\\
$26^{\circ}$ & 36.6-41.5 & $1.02\pm 0.06$ & $1.05\pm 0.09$ & $1.00\pm 0.08$ & $0.6\pm 0.1$ & $-$ & $0.24\pm 0.07 $ & $0.39\pm 0.06$ & $0.29\pm 0.04$ & $-$ &2.98\\
$42^{\circ}$ & 34.0-36.6 & $0.71\pm 0.06$ & $1.3\pm 0.2$ & $1.6\pm 0.2$ & $1.3\pm 0.1$ & $-$          & $-0.01\pm 0.08 $ & $0.22\pm 0.09$ & $0.11\pm 0.05$  & $-$ & 5.84\\
$42^{\circ}$ & 36.6-39.1 & $0.58\pm 0.07$ & $1.8\pm 0.2$ & $1.8\pm 0.3$ & $1.7\pm 0.3$ & $1.0\pm 0.3$ & $0.0\pm 0.1 $ & $0.3\pm 0.1$ & $0.20\pm 0.07$ & $-$ & 4.86\\ \hline
\end{tabular}
\end{table}
\endlandscape